\documentclass[aps,prc,onecolumn,showpacs,amsmath,amssymb,nofootinbib,11pt]{revtex4-2}
\usepackage{graphicx}
\usepackage{color}
\usepackage{times}
\usepackage{inputenc}
\usepackage{bm}
\usepackage{ulem}
\usepackage{multirow}
\usepackage{float}
\usepackage{url}
\usepackage{natbib}
\usepackage{mathrsfs}
\usepackage{physics}
\usepackage{comment}
\usepackage[table,xcdraw]{xcolor}
\usepackage{booktabs,makecell}
\usepackage{comment}
\usepackage[colorlinks=true,citecolor=blue,urlcolor=blue,linkcolor=blue]{hyperref}
\begin{document}
\title{Quarkyonic Neutron Stars as Candidates for the GW230529 Mass-Gap Object}

\author{Jeet Amrit Pattnaik}
\email{jeetamritboudh@gmail.com}
\affiliation{Department of Physics, Siksha ’O’ Anusandhan Deemed to be University, Jagamara, Bhubaneswar-751030, India}

\author{S. K. Patra}
\email{sureshpatra@soa.ac.in}
\affiliation{Department of Physics, Siksha ’O’ Anusandhan Deemed to be University, Jagamara, Bhubaneswar-751030, India}


\begin{abstract}
We examine whether quarkyonic equations of state (EOS) can account for compact objects in the $2.5$-$4.5\,M_\odot$ mass range reported for the GW230529 gravitational-wave event. The pressure-energy density and mass-radius (M-R) relations obtained from quarkyonic EOS models indicate a significant stiffening at high densities, allowing stable configurations beyond $2.5\,M_\odot$. The predicted M-R sequences extend into the GW230529 mass window while maintaining radii in the range of $\approx 13$-$15$~km, suggesting that quarkyonic stars can naturally populate the so-called compact object mass gap. These results imply that the heavier component of GW230529 could plausibly be a massive quarkyonic neutron star rather than a low-mass black hole.

\end{abstract}
\maketitle
%
\section{Introduction} \label{sec1}
The first direct detection of gravitational waves (GWs) from a black hole--black hole (BH-BH) merger in the landmark event GW150914 \cite{GW150914} not only inaugurated a new era of observational astrophysics but also provided a remarkable confirmation of Einstein's General Theory of Relativity \cite{GW150914}. Subsequently, the neutron star-neutron star (NS-NS) merger observed in GW170817 further enriched our understanding of compact objects and revealed, to some extent, the exotic nature of GW signals \cite{GW170817}. A particularly intriguing development followed with the multi-messenger event GW190814, arising from the merger of a $\sim 23\,M_{\odot}$ black hole with a mysterious secondary object of mass $2.50$-$2.67\,M_{\odot}$. This event posed more questions than answers regarding the true nature of the secondary component, which could be the lightest black hole, the heaviest neutron star, or possibly an entirely new class of exotic compact object.

According to current observations, the lightest known neutron star is HESS J1731-347, with an estimated mass of $M = 0.77^{+0.20}_{-0.17}\,M_{\odot}$ and a radius of $R = 10.4^{+0.86}_{-0.78}\,\mathrm{km}$ \cite{Doroshenoko2022}. In contrast, the heaviest confirmed neutron star has a mass of $M = 2.35^{+0.17}_{-0.17}\,M_{\odot}$ at the $1\sigma$ confidence level, as reported by Romani et al. \cite{Romani2022}. They also estimate the mass of PSR J1810+1744 to be $M = 2.13^{+0.04}_{-0.04}\,M_{\odot}$ \cite{Romani_2021}. It is important to note that these massive pulsars are rapidly rotating, and therefore their inferred masses exceed the maximum mass predicted by non-spinning Tolman-Oppenheimer-Volkoff (TOV) calculations, which may not be directly applicable in such cases.

Thus, the discovery of gravitational waves from compact binary mergers has opened a new avenue for exploring the physics of dense matter inside neutron stars. Recent detections by the LIGO–Virgo–KAGRA (LVK) collaboration have revealed objects whose inferred masses fall in the so–called mass gap between the heaviest observed neutron stars and the lightest black holes. Among these events, GW230529 is of particular interest because its primary component is estimated to lie within the $2.5$–$4.5\,M_\odot$ interval with $99 \%$ credibility, which has a collision with a neutron star of mass $M=1.2-2.0 M_{\odot}$ as the secondary object \cite{GW230529}.  The true nature of such an object remains uncertain: it could be an unusually massive neutron star supported by a stiff equation of state (EOS) or a low–mass black hole formed through an alternative evolutionary channel. The possibility of a black hole could be most unlikely, because the mass of the primary object is much less than the smallest possible black hole of mass $M=5 M_{\odot}$ at $99 \%$ credibility. 

Understanding whether matter at supranuclear densities can sustain stars above $2.5\,M_\odot$ requires reliable theoretical models of the EOS. 
At densities a few times the nuclear saturation density, baryonic degrees of freedom may coexist with deconfined quarks, leading to a gradual crossover rather than a sharp first–order transition \cite{McLerran2019,PhysRevD.102.023021,Dey2024}. This description, commonly referred to as quarkyonic matter, provides a unified framework in which both hadronic and quark components contribute to the pressure \cite{Dey2024}. In the present work, quarkyonic matter is implemented in a quarkyonic-inspired crossover framework, rather than as a sharp first-order phase transition. The formulation of quarkyonic matter is commonly understood as involving both nucleonic and quark degrees of freedom treated as effective quasi-particles, with a smooth crossover transition connecting the two regimes as the density increases. While nucleons remain the relevant effective degrees of freedom at low and intermediate densities and are governed by the same RMF interactions, the
crossover construction provides a physically motivated way to incorporate
additional pressure support from quark degrees of freedom at higher densities.
Because the crossover is implemented in a thermodynamically consistent manner,
the resulting high-density EOS should be regarded as an effective,
crossover-modified realization of quarkyonic-inspired stiffening, rather than as
an unchanged asymptotic quarkyonic phase. The crossover therefore does not
represent a sudden change in nucleon interactions, but rather a gradual
modification of the effective EOS as quark dynamics become relevant. The resulting EOS is typically stiffer than purely hadronic models at high density, allowing for larger maximum masses without violating causality.  In this work, we analyze the quarkyonic EOS within the relativistic mean–field (RMF) approach and examine the corresponding mass–radius (M–R) relations of compact stars. We aim to determine whether such EOS models can produce stable configurations within the GW230529 mass range. By comparing the predicted M–R curves with the observed mass window, we assess the possibility that quarkyonic stars can populate the compact object mass gap and thus provide a viable explanation for the massive component observed in GW230529.


\section{Theoretical Frameworks} \label{formalism}

At densities relevant to neutron star interiors, Quantum Chromodynamics (QCD) becomes highly non-perturbative, making ab-initio treatments extremely difficult. Lattice QCD, though successful at finite temperature and low baryon chemical potential, faces the sign problem at the high densities typical of compact stars, while perturbative QCD is quantitatively reliable only at asymptotically large chemical potentials well beyond the astrophysical domain. To bridge this intermediate regime, effective models are employed to capture the essential features of strong interactions. Among them, the relativistic mean-field (RMF) theory \cite{Dey2024} provides a practical and self-consistent framework in which the complex many-body problem is reduced to nucleons and quarks moving in classical meson mean fields that represent the average effect of interactions. This approach reproduces known saturation properties of nuclear matter and can be systematically extended to include additional degrees of freedom, such as quarkyonic components and dark matter \cite{Dey2024}, while ensuring thermodynamic consistency. The MIT bag model \cite{MIT1,MIT2} has also been widely used to describe deconfined quark matter by confining quarks within an effective bag that mimics the QCD vacuum pressure, providing a simple perturbative limit for comparison. Other frameworks addressing the non-perturbative region include chiral effective field theory at low density, Nambu–Jona-Lasinio \cite{NJL1,NJL2} and Polyakov–NJL \cite{PNJL} models for chiral symmetry restoration \cite{fischer2015}, Dyson–Schwinger \cite{DS} and functional-renormalization-group methods \cite{drews15} for continuum QCD studies, and hybrid or quark–hadron crossover constructions that interpolate between hadronic and quark equations of state. At the opposite extreme, perturbative QCD calculations provide constraints at ultra-high density where the coupling becomes weak. The mean-field description thus serves as a reliable and computationally tractable link between these two regimes, allowing consistent exploration of dense QCD matter in the density range probed by neutron stars.

The recently proposed quarkyonic model of L. McLerran and S. Reddy \cite{McLerran2019}, which is developed by T. Zhao and J. M. Lattimer \cite{PhysRevD.102.023021}, considering the $\beta-$equalibrium condition in the neutron star matter is quite successful to describe the masses of dense objects, like neutron stars etc. Further, this model extended to Dark Matter (DM) sector with an Effective Relativistic Mean Field (E-RMF) frame-work \cite{ERMF7,ERMF9, patt21, patt22} of hadronic medium in the periphery of the Quark core covered with the BPS equation of state \cite{BPS} on the crust of the neutron star. In this direction, the readers are also advised to see our recent study with the two-fluid approach on dark matter admixed quarkyonic stars \cite{patt25}. In this present work, we have used this model without taking DM into account in the medium of the NS matter. The IOPB-I \cite{IOPB} and G3 \cite{G3} parametrizations are used in the calculations. (We refer the readers to Refs. \cite{IOPB,G3,patt21,patt22,Dey2024} for E-RMF theory and the detailed procedure of the formalism to evaluate the EOS for the quarkyonic neutron star models.) In brief, the transition density from Quarkyonic to hadronic phase is taken as $n_t=0.3,0.4,0.5$ fm$^{-3}$ with the confinement scale $\Lambda_{cs}$ = 800 and 1400 MeV. The parameter $n_t$ characterizes the central density of the crossover region and
does not represent a sharp phase boundary.
 \\

The quarkyonic equations of state employed in this work are constructed within a smooth crossover framework, following the interpolated EOS prescription introduced by Masuda \emph{et al.}~\cite{Masuda2013} and subsequently implemented by Han \emph{et al.}~\cite{sophia_han_19}.
This approach differs from a sharp hybrid (Maxwell or Gibbs) construction \cite{patra07a, patra07b} and instead describes a continuous
transition between hadronic and quark (or quarkyonic-inspired) degrees of
freedom. In this approach, the pressure is smoothly
interpolated between hadronic and quarkyonic-inspired sectors using a continuous
switching function characterized by a finite width parameter $\Gamma$, which
controls the extent of the crossover region. This construction avoids unphysical
discontinuities and ensures numerical stability, while preserving the essential
high-density stiffening associated with quarkyonic dynamics.

In this framework, the transition density $n_t$ represents a characteristic
onset scale for quarkyonic degrees of freedom, while the crossover itself extends
over a finite density interval controlled by a width parameter $\Gamma$.
Consequently, the EOS is not required to coincide exactly with the purely
nucleonic EOS up to a single sharp density threshold.

The crossover is implemented through an interpolation of the pressure in baryon
density space according to ~\cite{sophia_han_19}
\begin{equation}
P(n) = P_{\rm had}(n)\,f_{-}(n) + P_{\rm qk}(n)\,f_{+}(n).
\label{eq:Pinterp}
\end{equation}
Here $P_{\rm had}(n)$ and $P_{\rm qk}(n)$ denote the purely nucleonic and
quarkyonic-inspired pressures, respectively. The smooth switching functions
satisfy
\begin{equation}
f_{-}(n) + f_{+}(n) = 1,
\qquad
f_{\pm}(n) =
\frac{1}{2}
\left[
1 \pm \tanh\!\left(\frac{n-n_t}{\Gamma}\right)
\right].
\label{eq:weights}
\end{equation}
The parameter $\Gamma$
controls the width of the crossover region and is treated phenomenologically to
ensure smoothness and numerical stability.


As emphasized in Refs. ~\cite{Masuda2013, sophia_han_19}, pressure interpolation requires a thermodynamically consistent reconstruction of the energy density. Accordingly, the energy density in the crossover region is given by
\begin{equation}
\varepsilon(n)
=
\varepsilon_{\rm had}(n)\,f_{-}(n)
+
\varepsilon_{\rm qk}(n)\,f_{+}(n)
+
\Delta\varepsilon(n),
\label{eq:energy}
\end{equation}
which corresponds to Eq.~(21) of Ref.~\cite{sophia_han_19}. The additional term
$\Delta\varepsilon(n)$ arises from the density dependence of the switching
function and is given explicitly by
\begin{equation}
\Delta\varepsilon(n)
=
n \int_{n_t}^{n}
dn'\,
\frac{\varepsilon_{\rm had}(n') - \varepsilon_{\rm qk}(n')}
{n'}\, g(n'),
\label{eq:deltaeps}
\end{equation}
with
\begin{equation}
g(n') = \frac{d f_{+}(n')}{dn'}
      = \frac{1}{2\Gamma}
        \mathrm{sech}^2\!\left(\frac{n'-n_t}{\Gamma}\right).
\end{equation}
This additional contribution is a direct consequence of enforcing the
thermodynamic identity
\begin{equation}
P = n^2 \frac{\partial}{\partial n}\left(\frac{\varepsilon}{n}\right),
\end{equation}
and is not an ad hoc modification. Because the interpolation modifies the mapping between baryon density $n$, energy
density $\varepsilon$, and pressure $P$, equations of state that are nearly
identical in the $P(n)$ representation may appear separated when expressed in the $P(\varepsilon)$. The apparent separation of EOS curves at low energy
densities therefore reflects the smooth crossover nature of the interpolation
rather than a modification of the underlying nucleonic RMF dynamics. We
emphasize that the nucleonic parameter sets (G3 and IOPB-I) are kept fixed
throughout and are not refitted for different choices of $n_t$.

\section{Results and Discussions} \label{result}
The recently computed Quarkyonic Model of Dey et al \cite{Dey2024} is used to calculate the Mass (M), Radius (R), compactness (C) and dimensionless tidal deformability ($\Lambda$) for an isolated non-spinning binary neutron star. The hybrid quarkyonic model of equation of state is used to solve the TOV equations for the evaluation of M, R and $\Lambda$. The results are presented and discussed in the following subsequent subsections. 

\begin{figure*}
\includegraphics[width=1.0\columnwidth]{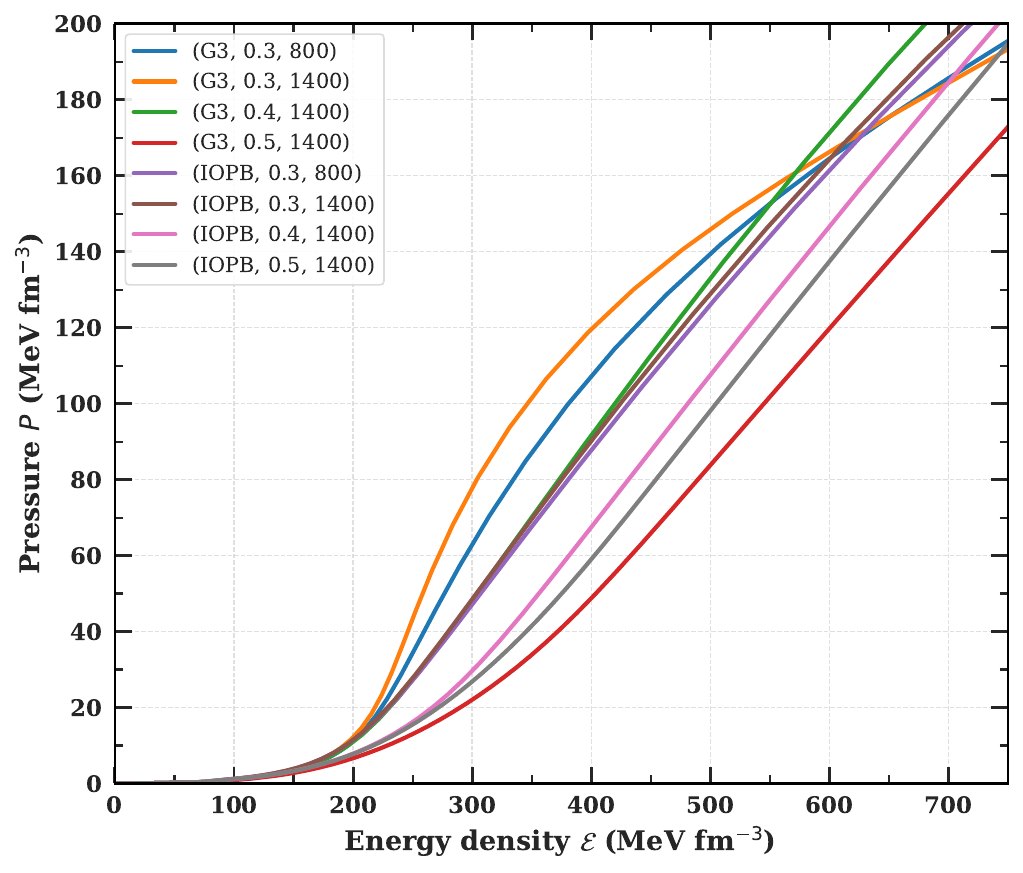}
\caption{Pressure as a function of energy density for the quarkyonic equations of state at a transition density 
$n_t = 0.3,0.4,0.5~\mathrm{fm^{-3}}$ with confinement scales 
$\Lambda_{\mathrm{cs}} = 800$ and $1400$~MeV for the G3 and IOPB models. }
\label{fig:eos}
\end{figure*}

We note that the equations of state shown in Figure~\ref{fig:eos} are presented in the pressure--energy density representation, $P(\varepsilon)$. In the crossover framework adopted here, the interpolation is performed in baryon density space, $P(n)$, while the energy density is obtained from Eq.~(\ref{eq:energy}). As a result, equations of state that are nearly identical in the $P(n)$ representation may appear separated when expressed as $P(\varepsilon)$, particularly in the vicinity of the crossover region. This behavior is a generic feature of smooth crossover constructions and does not imply a modification of the underlying nucleonic RMF dynamics.

In Figure~\ref{fig:eos}, it is noticed that the purely hadronic IOPB-I parameterization is stiffer than G3 in the low-to-moderate density regime; the addition of quarkyonic degrees of freedom alters the stiffness hierarchy at higher densities. Specifically, the quarkyonic variants order as $(\mathrm{Force},\,n_t,\, \Lambda_{cs})=$ $(\mathrm{G3},\,0.3,\,1400) \succ (\mathrm{G3},\,0.3,\,800) \succ (\mathrm{IOPB},\,0.3,\,800)$ in terms of pressure at fixed energy density. In other words, the G3-based EOS with the larger confinement scale attains the highest pressure among the set, followed by the G3 case with the smaller confinement scale, and finally the IOPB-I quarkyonic curve.
This reordering can be understood as the result of two competing effects. 
First, the baseline stiffness of the baryonic model determines how much additional pressure the quarkyonic sector can supply: a relatively soft baryonic foundation (G3) admits a larger fractional stiffening once quark momentum modes and corresponding repulsive components are introduced. 
Second, the confinement scale $\Lambda_{\mathrm{cs}}$ controls the onset and strength of quark contributions; a larger $\Lambda_{\mathrm{cs}}$ effectively delays deconfinement and enhances the high-momentum pressure contribution, producing a stronger net stiffening for the same transition density $n_t=0.3\,\mathrm{fm^{-3}}$. Combined, these effects explain why the G3 family, when supplemented with a high confinement scale, outpaces the IOPB-I quarkyonic curve despite IOPB-I being the stiffer baryonic model originally. A similar type of behaviour can be observed for other reported quarkyonic cases.

\begin{table}[htbp]
\centering
\caption{The maximum masses obtained from different combinations of (Force, $n_t$, $\Lambda_{cs}$) and their corresponding properties of the quarkyonic neutron stars, such as maximum radius $R_{max}$, compactness $C_{max}$ and dimensionless tidal deformability $\Lambda_{max}$. The canonical deformability $\Lambda_{1.4}$ is also given for reference.}
\label{tab1}
\begin{tabular}{lccccc}
\hline
\hline
\textbf{($\mathrm{Force},\, n_t,\, \Lambda_{cs}$)} & \textbf{$M_{\mathrm{max}}$ ($M_{\odot}$)} & $R_{\mathrm{max}} (km)$ & \textbf{$C_{\mathrm{max}}$} & \textbf{$\Lambda_{\mathrm{max}}$} & \textbf{$\Lambda_{1.4}$} \\
\hline
(G3, 0.3, 800)    & 2.75 & 14.54 & 0.279 & 16.16 & 957.12  \\
(G3, 0.3, 1400)   & 2.95 & 15.16 & 0.287 & 13.31 & 1187.23 \\
(G3, 0.4, 1400)   & 2.56 & 13.61 & 0.277 & 14.58 & 1210.00 \\
(G3, 0.5, 1400)   & 1.98 & 11.33 & 0.257 & 21.78 & 533.29  \\
(IOPB, 0.3, 800)  & 2.50 & 13.65 & 0.270 & 12.61 & 626.95  \\
(IOPB, 0.3, 1400) & 2.54 & 13.76 & 0.272 & 12.08 & 581.77  \\
(IOPB, 0.4, 1400) & 2.24 & 12.30 & 0.268 & 13.08 & 494.87  \\
(IOPB, 0.5, 1400) & 2.15 & 11.93 & 0.265 & 13.66 & 448.42  \\
\hline
\hline
\end{tabular}
\end{table}

The figure also shows that the quarkyonic crossover remains smooth in all cases: there is no abrupt plateau or discontinuity in $P(\varepsilon)$, and the slopes remain positive. Practically, the elevated pressure in the $(\mathrm{G3},\,0.3,\,1400)$ model implies the largest increase in the maximum mass and a modest inflation of radii at fixed mass relative to the other curves; the intermediate $(\mathrm{G3},\,0.3,\,800)$ case produces a smaller but still significant shift, while the $(\mathrm{IOPB},\,0.3,\,800)$ curve remains closest to the purely baryonic IOPB-I behavior at the densities shown.
\begin{figure*}
\includegraphics[width=1.0\columnwidth]{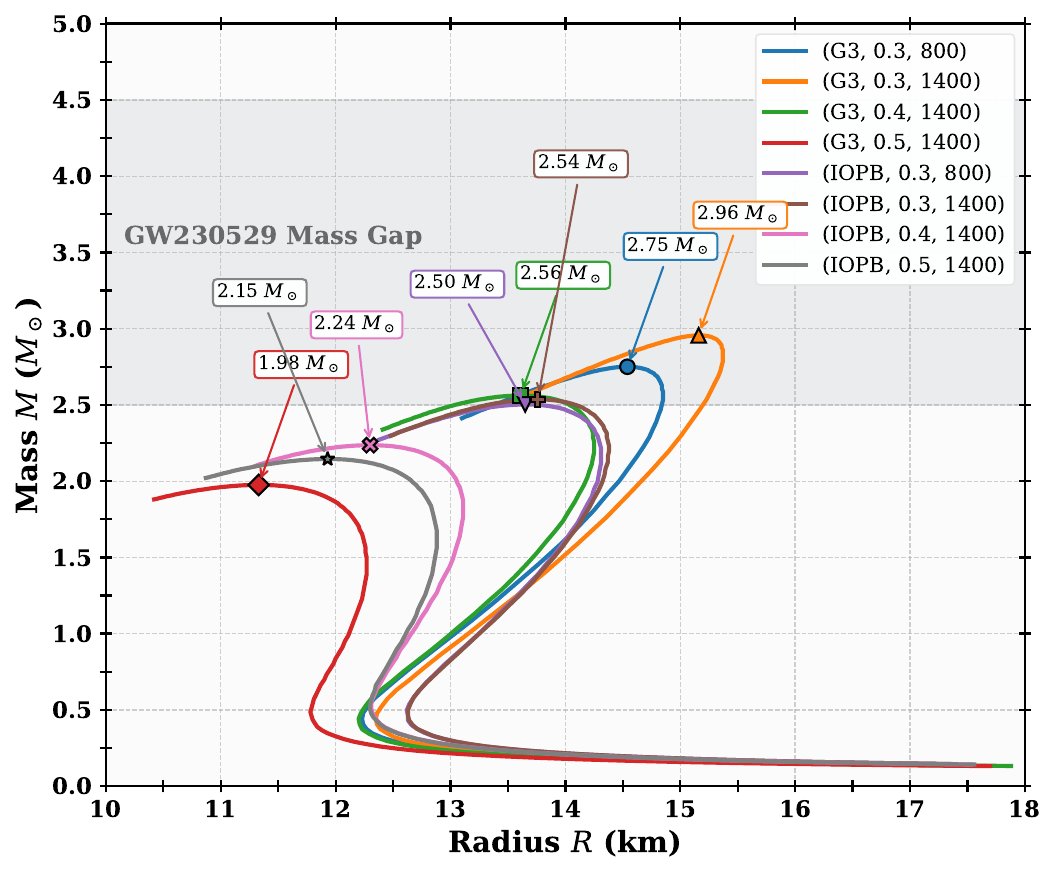}
\caption{M-R relations for the EoSs shown in Fig. \ref{fig:eos}, along with the GW230529 observational data \cite{GW230529}.}
\label{fig:mr}
\end{figure*}

The corresponding mass-radius ($M$-$R$) characteristics derived from the quarkyonic equations of state are shown in Fig.~\ref{fig:mr}. 
Each curve represents the stellar sequence obtained from integrating the Tolman-Oppenheimer-Volkoff equations for the specified EOS parameters. The horizontal shaded band denotes the $2.5$-$4.5\,M_\odot$ interval associated with the heavier component of the GW230529 event, which defines the observationally inferred ``mass gap'' between typical neutron stars and low-mass black holes. It is important to emphasize that, within the present crossover framework, the
maximum mass of a neutron star is governed by the effective stiffness of the
equation of state at supranuclear densities. While enforcing thermodynamic
consistency in a smooth crossover construction induces a modification of the
energy density, as discussed by Masuda \emph{et al.} \cite{Masuda2013} and implemented by Han
\emph{et al.} \cite{sophia_han_19}, this does not eliminate the dominant role of quarkyonic-inspired
dynamics in providing enhanced pressure support at high densities.
Accordingly, the large maximum masses obtained in this work reflect a robust
consequence of incorporating quarkyonic-inspired stiffening within a
thermodynamically consistent EOS, rather than a fine-tuned artifact of the
interpolation procedure. Interpolation-induced variations at lower densities
primarily influence the stellar radius and have a subleading impact on the
stability threshold.

Among the models examined, the $(\mathrm{G3},\,0.3,\,1400)$ case produces the stiffest sequence, yielding the largest radii and the highest maximum mass, which slightly exceeds $2.9\,M_\odot$. The $(\mathrm{G3},\,0.3,\,800)$ model lies below it, while the $(\mathrm{IOPB},\,0.3,\,800)$ curve remains the softest and terminates near $2.75\,M_\odot$. This ordering mirrors the relative stiffness identified in the EOS plot: a higher confinement scale strengthens the quark pressure and extends the stable branch to larger masses, whereas the stiffer baryonic base (IOPB-I) contributes less relative stiffening once quarkyonic corrections are added. The resulting inversion in stiffness hierarchy demonstrates that the quarkyonic contribution affects 
each hadronic framework differently, depending on the balance between nucleonic and quark pressures at the transition density. Although, the IOPB-I equation of state at (IOPB, 0.3, 800) crosses the G3-EOSs at energy density $\epsilon\sim 640$ MeVfm$^{-3}$, it is not sufficient enough to fill up the mass gap in the range $M=2.5-4.5 M_{\odot}$. It just touches the lower limit of the mass range for the primary object estimated in the GW230529 observation. Further, the analysis of the M-R relation with transition density $n_t=0.5$ fm$^{-3}$ is performed and the mass of the neutron star found to be lower than 2.5 $M_{\odot}$ (See Fig.~\ref{fig:mr}) with both the considered parameter sets (IOPB-I and G3, See Fig.~\ref{fig:mr}). This gives an indication that the transition of hadron to quark phase for the primary object in the GW230529 event is possible in the density region $n_t\sim$ 0.3 to 0.4 fm$^{-3}$. As we have shown in Table~\ref{tab1}, the stellar properties vary systematically with $\Lambda_{\mathrm{cs}}$ and $n_t$.

Among the considered cases, five sequences intersect the GW230529 mass window, confirming that the quarkyonic EOSs considered here can produce stable, non-rotating configurations within the observed range. The predicted radii for these high-mass stars fall between roughly $13$ and $15$~km, indicating compact but non-collapsed structures consistent with current gravitational-wave constraints. No turning points or instabilities appear up to the maximum masses shown, signifying that the stellar configurations remain stable throughout the quarkyonic regime. The smooth progression of the $M$-$R$ curves also verifies that the crossover treatment preserves causality and avoids any discontinuous transition that might lead to twin-star behavior.

The compactness parameter $C_{max}=M_{max}/R_{max}$ of the hybrid star for different combinations of parameter set, transition density and confinement scale (Force, $n_t$, $\Lambda_{cs}$ of the quarkyonic star are listed in Table 1. From the table, it is evident that the canonical tidal deformability $\Lambda_{1.4}$ is maximum for (G3, 0.4, 1400) case with a maximum mass of $M=2.56 M_{\odot}$ and it is minimum for (IOPB, 0.5, 1400) which has a maximum mass of $M=2.15 M_{\odot}$. Thus, in the case of a quarkyonic star, the canonical deformability does not follow any pattern with the mass of the neutron star. 

\section{Summary and Conclusions}
\label{sec:conclusion}
In summary, the results presented in this work should be interpreted within a
smooth, thermodynamically consistent quarkyonic-inspired crossover framework,
rather than within a sharp hybrid equation-of-state picture. We have formulated
the equation of state developed by Dey \emph{et al.} within this crossover
approach, where the parameter $n_t$ characterizes the central density of the
crossover region associated with the onset of quarkyonic-inspired dynamics, and
does not represent a sharp phase boundary. In the present study, we consider
representative crossover densities $n_t = 0.3$ and $0.4~\mathrm{fm}^{-3}$. In other words, for the crossover densities $n_t = 0.3$-$0.4~\mathrm{fm}^{-3}$
considered in this work, quark degrees of freedom become increasingly relevant in the inner core of the neutron star and contribute significantly to the pressure support, while the matter is described within an effective,
crossover-modified equation of state rather than as a pure quark phase. Within
this framework, nucleonic degrees of freedom govern the low- and
intermediate-density regions, while quarkyonic-inspired contributions gradually
enhance the pressure support at supranuclear densities beyond $n_t$. The resulting equations of state, when employed as input to the
Tolman-Oppenheimer-Volkoff (TOV) equations yield neutron star configurations with large maximum masses, sizable radii, and consistent tidal deformabilities, in agreement with current observational constraints. Our results demonstrate that quarkyonic-inspired stiffening within a crossover-modified equation of state can support compact stars with masses exceeding $2.5\,M_{\odot}$ for the range of crossover densities explored in this work. In this context, the primary object associated with GW230529 is consistent with an interpretation as a massive neutron star, possibly a quarkyonic-inspired neutron star within the present crossover framework, rather than a low-mass black hole.

\bibliographystyle{apsrev4-2}
\bibliography{nstar}

\end{document}